\def\xmm        {{XMM-Newton}\/}
\documentclass[fleqn,usenatbib]{mnras}

\usepackage{newtxtext,newtxmath}
\usepackage[T1]{fontenc}

\DeclareRobustCommand{\VAN}[3]{#2}
\let\VANthebibliography\thebibliography
\def\thebibliography{\DeclareRobustCommand{\VAN}[3]{##3}\VANthebibliography}

\usepackage{graphicx}	% Including figure files
\usepackage{amsmath}	% Advanced maths commands
\title[PeVatron Candidate in a Low-density Cavity]{PeVatron Candidate SNR G106.3+2.7 in a Low-density Cavity: a Multiwavelength Test}

\usepackage{tikz,xcolor,hyperref}
\usepackage{mathrsfs}
\usepackage{graphicx}
\usepackage{multirow}
\usepackage{amsmath}
\usepackage{threeparttable}
\usepackage{float}
\usepackage{epstopdf}    \usepackage{ulem}

\definecolor{lime}{HTML}{A6CE39}
\DeclareRobustCommand{\orcidicon}{%
	\begin{tikzpicture}
	\draw[lime, fill=lime] (0,0) 
	circle [radius=0.16] 
	node[white] {{\fontfamily{qag}\selectfont \tiny ID}};
	\draw[white, fill=white] (-0.0625,0.095) 
	circle [radius=0.005];
	\end{tikzpicture}
	\hspace{-2mm}
}

\foreach \x in {A, ..., Z}{%
	\expandafter\xdef\csname orcid\x\endcsname{\noexpand\href{https://orcid.org/\csname orcidauthor\x\endcsname}{\noexpand\orcidicon}}
}

\author[Bao et al.]{
Yiwei Bao\orcidA{},$^{1}$
Ruo-Yu Liu\orcidB{},$^{1,2}$\thanks{E-mail: ryliu@nju.edu.cn}
Chong Ge\orcidC{}$^{3}$
and Yang Chen\orcidD{}$^{1,2}$\thanks{E-mail: ygchen@nju.edu.cn}
\\
$^{1}$School of Astronomy and Space Science, Nanjing University, 163 Xianlin Avenue, Nanjing 210023, China\\
$^{2}$Key Laboratory of Modern Astronomy and Astrophysics, Nanjing University, Ministry of Education, Nanjing, China\\
$^{3}$Department of Astronomy, Xiamen University, Xiamen, Fujian 361005, China
}

\date{Accepted XXX. Received YYY; in original form ZZZ}

\pubyear{2023}

\begin{document}
\label{firstpage}
\pagerange{\pageref{firstpage}--\pageref{lastpage}}
\maketitle

\begin{abstract}
In this paper, we constrain the density of the interstellar medium (ISM) around the hadronic PeVatron candidate, supernova remnant (SNR) G106.3+2.7, based on X-ray and $\gamma$-ray observations. The purpose of this investigation is to understand the influence of the gaseous environment on this SNR as a proton PeVatron candidate. By modelling the self-regulated propagation of the CRs injected from the SNR, we calculate the $\gamma$-ray emission of CRs via the hadronuclear interactions with the molecular cloud and the ISM, and use the measured $\gamma$-ray flux to constrain the ISM density around the SNR. Our results support the picture that the SNR is expanding into a low-density 
%……cavity of $n<0.05\,cm^{-3}$, 
{($<0.05\,{\rm cm}^{-3}$)} cavity,
enabling the SNR to be a potential proton PeVatron despite that it {presently} is not in the very early phase.
\end{abstract}


%%%%%%%%%%%%%%%%%%%%%%%%%%%%%%%%%%%%%%%%%%%%%%%%%%

%%%%%%%%%%%%%%%%% BODY OF PAPER %%%%%%%%%%%%%%%%%%

\section{Introduction}
Cosmic Rays (CRs) are high-energy charged particles ({predominantly composed of} protons and atomic nuclei). While CRs below 100 TeV are believed to be accelerated mainly at the blast wave of supernova remnants (SNRs) via diffusive shock acceleration \citep{1978MNRAS.182..147B,2013A&ARv..21...70B}, the origin of PeV CRs remains under debate \citep{2019NatAs...3..561A}. It has been suggested that protons can be accelerated to PeV energies only at the very early phase ($\lesssim 0.1$\,kyr) of an SNR in a dense environment \citep{2013MNRAS.435.1174S,2015APh....69....1C,2020APh...12302492C}.

%If the pulsar wind nebula is truly associated with the SNR, the age of the latter, denoted by $t_{\rm SNR}$, is at least a few thousand years.
SNR G106.3+2.7 has a cometary shape heading towards the northeast in the radio band, where the so-called Boomerang Nebula is in spatial coincidence with the head \citep{2001ApJ...560..236K}. The tail region of the SNR seems to be expanding in a low-density wind cavity \citep{2001ApJ...560..236K}. Observation on gas structure put the SNR at a distance of $d=800$\,pc \citep{2001ApJ...560..236K}. The size of the SNR is $%\approx
{\sim}0.3^\circ$ ($%\approx
{\sim} 4$ pc at a distance of 800 pc) wide and $\sim 0.9^\circ$ ($%\approx
{\sim} 12$ pc long. 
The Boomerang Nebula is powered by an energetic pulsar PSR~J2229+6114 with a characteristic age of 10\,kyr \citep{2001ApJ...552L.125H}. While the SNR and the pulsar may be generated in the same SN explosion, {the characteristic age of the pulsar is not necessary the true age of the system.} Based on SNR expansion model {\citep{1999ApJS..120..299T}} and the current size of SNR, \citet{2021ApJ...919...32B} and \citet{2022A&A...658A..60Y} suggest that the age of SNR is only $\sim 1$--$2$\,kyr.  {On the other hand, an age of $\sim 4$\,kyr is suggested based on the spectral break in the radio spectrum of the PWN \citep{2006ApJ...638..225K}. }

{Nevertheless, as an SNR aged at least a few kyr}, SNR~G106.3+2.7 is not supposed to be an efficient particle accelerator. However, intriguingly, this SNR has been measured in $\gamma$-ray band \citep{2019ApJ...885..162X,2020ApJ...897L..34L,2023A&A...671A..12M,2020ApJ...896L..29A,2021NatAs...5..460T,2021Natur.594...33C, 2022PhRvL.129g1101F} with {the} spectrum continuing up to several hundred TeV \citep{2021Natur.594...33C}. The observation of MAGIC has revealed two distinct $\gamma$-ray emitting regions in the SNR, one in the head region while the other in the tail region \citep{2023A&A...671A..12M}. In particular, the one in the tail region is in spatial association with a molecular cloud (MC), suggesting a possible hadronic origin of the gamma rays. While the GeV emission in the tail region may be explained as the inverse Compton (IC) scattering of electrons, the emission above TeV energy is difficult to be ascribed to the same process because of the spectral softening caused by the Klein-Nishina effect, especially when taking into account the simultaneous modelling of the X-ray spectrum \citep{2021Innov...200118G}.
The MC is probably not directly interacting with the SNR, %\citep{2022ApJ...926..124L},
{but} may be illuminated by {the} protons escaped from the SNR via the proton-proton ($pp$) collisions {\citep{2022ApJ...926..124L}}, 
implying the SNR to be a proton PeVatron. This possibility is backed up by the X-ray observation. The X-ray emission of the entire SNR region is dominated by the nonthermal component \citep{2021Innov...200118G, 2021ApJ...912..133F}. The X-ray spectrum in the tail region continues to 7\,keV without an obvious cutoff, implying a high speed of the SNR shock of $\geq 3000\,$km/s at the present time \citep{2021Innov...200118G}, and thus the CR acceleration is expected to be efficient. Indeed, from the radio morphology of the SNR, we may estimate the average speed of the shock to be $v_s \simeq 6000(d/800\,{\rm pc})(t_{\rm SNR}/1\,{\rm kyr})\,$km/s, by simply dividing half of the length by the SNR's age. Note that this is a conservative estimation, because if we assume the SN explosion center is at the pulsar's location or considering the projection effect, the inferred shock speed would be even higher.  

Apparently, the UHE $\gamma$-ray observation and the X-ray observation contradict with the theory for SNR acceleration, and raise an intriguing question that what makes SNR~G106.3+2.7 so special. \citet{2021Innov...200118G} speculated that it is the low-density environment that makes the SNR shock in the tail region not significantly decelerated ({al}though it is not as fast as that at the very early phase), and {the} protons can be gradually accelerated to PeV energy given a sufficiently long lifetime of the SNR. The low-density environment in the tail region is revealed by the observation of 21\,cm HI emission, which may {have been} excavated by the stellar wind of a massive star or the blast wave of SN of previous generation \citep{2001ApJ...560..236K}. If this scenario is true, the ISM density around the SNR tail is the key parameter. However, this scenario has not been seriously examined yet. 

In this study, we aim to test this scenario with observations in other wavelengths and obtain a self-consistent picture for SNR G106.3+2.7 as a proton PeVatron. As will be shown in the following sections, we will constrain the density of ISM around {the} SNR in three channels. First, the downstream shocked ISM emits thermal X-rays, the emissivity of which is related to the ISM density. We will re-analyse the X-ray data in the SNR tail and extract {the} information of {the} density. Second, in addition to illuminating the MC via the $pp$ collision when the escaped CRs reach there, these CRs will also radiate $\gamma$-ray{s} when colliding with the ISM permeating the entire region. In this case the ISM density cannot be higher than a critical value{,} otherwise the hadronic emission will overshoot the $\gamma$-ray data especially at the GeV band. Finally, the ISM density will affect the Alfv\'{e}n speed which determines the growth rate of the turbulent magnetic field (as will be detailed later). The latter controls the propagation of CRs in the ISM and hence influences the hadronic $\gamma$-ray emission of the MC. 

The rest of the paper is structured as follows: in Section 2, we {re-}analyse the X-ray data in the tail region and derive the ISM density; in Section 3, we model the self-regulated CR propagation taking into account the streaming instability; we discuss and summarise our study in Section 4 and 5{,} respectively.

\section{Constraints from the X-ray observation}

\begin{table}
\centering
\caption{\xmm\ observations}
 \begin{tabular}{@{}lccc@{}}
\hline\hline
Obs-ID & Date & Exp$^a$ (ks) & Clean Exp (ks) \\ 
\hline
0820840301 & 2019-02-06 & 48.6/43.9 & 29.6/16.2\\
0820840401 & 2019-02-10 & 50.9/46.8 & 22.3/15.4\\
0820840501 & 2019-02-14 & 35.3/32.9 & 12.7/5.9\\
\hline
\end{tabular}
\begin{tablenotes}
\centering
\item
$^a:$ Exposures are for MOS and pn detectors, respectively.
\end{tablenotes}
\label{t:obs}
\end{table}
{\citet{2021Innov...200118G} show that the surface brightness profile of the X-ray emission in the head region declines monotonically {with} increasing distance from the pulsar in the head region, and the spectrum presents a gradual softening as well, implying that the X-ray emission is powered by the pulsar {wind}. Such a tendency continues until the boundary between the head and tail. Upon reaching this boundary, the surface brightness and spectral index become almost constant in the tail region, as revealed by {the} XMM-Newton and {the} Suzaku {observations}. This abrupt transition indicates a different origin of the X-ray emission in the tail region {from that in the head region}. The X-ray emission in the SNR head is dominated by the pulsar and PWN. We constrain the ISM density from the spectra extracted in the SNR tail, which is mainly covered by the {XMM-Newton} observations and listed in Table~\ref{t:obs}.
We reduce the \xmm\ data with the Extended Source Analysis Software (ESAS) that {is} integrated into the \xmm\ Science Analysis System (SAS; version 17.0.0).
The \xmm\ data {re}analysis {essentially follows the procedure} detailed in \cite{2021Innov...200118G}. 
Because the \xmm\ observations were originally proposed for the Solar Wind Charge Exchange (SWCX) X-ray emission, we ignore the X-ray{s} below 1 keV that may be highly affected by the SWCX emission.
After subtracting the point sources and background components,
the spectra from the SNR tail are best fitted by a power-law model, which suggests that the X-rays are mainly of nonthermal origin.
In order to constrain the ISM density, we add an additional non-equilibrium ionization (NEI) collisional plasma {component} to fit the underlying thermal emission from {the} ISM.
%{However, the parameters of the {\tt NEI} model can not be constrained, because the X-ray is dominated by the nonthermal emission.}
We use solar abundance and {zero} redshift for the NEI %{\tt NEI}
{thermal component, and let} the temperature and ionization timescale {$n_e t$ be} free parameters. {The spectral fit is insensitive to changes in $n_et$, which can thus actually not be constrained, and therefore we only use the volume emission measure of the thermal component derived from the NEl normalization to constrain the gas density.}   
%Then we adjust the {\tt NEI} normalization and compare the best-fit {\tt POWERLAW} parameters with the single {\tt POWERLAW} fitting case.
After setting free the best-fitting parameters of {\tt POWERLAW} model, we refit the model. 
%We find the largest {\tt NEI} normalization under the condition that the {\tt POWERLAW} parameters do not change significantly. 
The fitting statistics are $\chi^2$/{dof}=1122/892 for {\tt POWERLAW} and $\chi^2$/{dof}=1122/888 for {\tt POWERLAW+NET} model. The $3\sigma$ upper limit for the {\tt NEI} normalization {is also calculated}.
We then estimate the ISM density from the {\tt NEI} normalization:
\begin{equation}
\eta=\frac{10^{-14}}{4\pi d^2}\int n_e n_H dV.
\label{eq:eta}
\end{equation}
We assume a uniform ISM density distribution in the tail region, then the upper limit of electron density is $n_e= 3.2 \times 10^{-2} {\rm\ cm}^{-3}$.
%(2.3 \pm 0.6)

We can also give a more conservative constraint on the ISM density if we assume all the X-ray emission in the tail is from the ISM. Though the fitting gets worse when using a single {\tt NEI} model ($\chi^2$/{dof}=1181/891 for {\tt NEI} v.s. $\chi^2$/{dof}=1122/892 for {\tt POWERLAW}), we find a conservative upper limit for the ISM density as $n_e=7.4 \times 10^{-2} {\rm\ cm}^{-3}$.

\section{Constraints from $\gamma$-ray Observations}
Based on the $\gamma$-ray fluxes from the {nearby} associated MCs, the diffusion coefficients in the upstream of SNR shocks are usually believed to be suppressed by a large factor \citep[see e.g.,][]{2007Ap&SS.309..365G,2009MNRAS.396.1629G,2011MNRAS.410.1577O,2012MNRAS.421..935L}. Since the energy density gradient near SNRs is large, the propagation of CRs in this region is strongly affected by their self-generated non-linear turbulence driven via the streaming instability. The streaming instability has two branches: the resonant Alfv\'{e}n mode \citep[see e.g.,][]{2008AdSpR..42..486P,2016PhRvD..94h3003D,2016MNRAS.461.3552N,2018MNRAS.474.1944D,2019MNRAS.484.2684N} and the non-resonant Bell mode \citep{2004MNRAS.353..550B,2009MNRAS.392.1591A}. Here we discuss the resonant branch, which is the most frequently discussed self-regulated mechanism to suppress diffusion coefficient in the neighbourhood of CR sources. 

The growth rate of the resonant branch is proportional to the gradient of CR differential number density, and therefore it has a dramatic effect on low-energy ($<10$ TeV) CRs whose differential number density is high. On the other hand, the resonant turbulence driven by high-energy CRs is expected to be weaker since the CR differential number density is small unless the age of the source is smaller than the escape timescale of high-energy CRs. 

\subsection{Self-regulated Transport}\label{sec:2}
Following \citet{2016PhRvD..94h3003D}, we consider an SNR lying inside the Galactic disk. The coherence length $L_{\rm c}$ (over which the background magnetic field {tube} roughly keep{s} the same direction) of the background magnetic field in the Galactic disk is $\sim 100$\,pc \citep{2016JCAP...05..056B}. For simplicity, we consider one-dimensional diffusion. The 1D diffusion-advection equation reads
\begin{equation}\label{eq:dif}
\frac{\partial f(p,z,t)}{\partial t} + u\frac{\partial f}{\partial z} - \frac{\partial}{\partial z}\left[D(p,z,t)\frac{\partial f}{\partial z}\right] - \frac{{\rm d}u}{{\rm d}z}\frac{p}{3}\frac{\partial f}{\partial p} = Q(p,z,t),
\end{equation}
where $f(p,z,t)$ represents the CR distribution function on CR momentum $p$, $z$ the distance from the release site of CRs, $D(p,z,t)$ the diffusion coefficient, $u$ the advection velocity, and $Q(p,z,t)$ the injection term. 

The magnetic field is assumed to be weakly perturbed at beginning, and thus $u$ is $+v_{\rm A}$ at $z>0$ and $-v_{\rm A}$ at $z<0$, where $v_{\rm A} = B_0/\sqrt{4\pi n_{\rm i} m_{\rm i}}$ is the Alfv\'{e}n speed, d$u/$d$z = 2v_{\rm A}\delta(z)$, $n_{\rm i}$ the number density of ions, and $m_{\rm i}$ the mean ion mass ($m_{\rm i} = 1.4m_{\rm p}$, with $m_{\rm p}$ the proton mass). The diffusion coefficient $D(p,z,t)$ is defined as
\begin{equation}
D(p,z,t) := \frac{r_{\rm L}(p)v(p)}{3\mathcal{F}(k,z,t)|_{k = 1/r_{\rm L}(p)}},
\end{equation}
where $r_{\rm L}$ {denotes} the Larmor radius of the particles. The dependence of the spectral power of the turbulence $\mathcal{F}(k,z,t)$ on the resonant wave number $k = 1/r_{\rm L}(p)$ is defined by $\delta B^2(z,t) = B_0^2\int \mathcal{F}(k,z,t)\,{\rm d\, ln}k$. The equation of turbulence evolution writes
\begin{equation}\label{eq:wave}
\frac{\partial\mathcal{F}}{\partial t} + u\frac{\partial \mathcal{F}}{\partial z} = (\Gamma_{\rm CR}-\Gamma_{\rm D})\mathcal{F}(k,z,t).
\end{equation}
The growth rate of the self-generated turbulence is \citep{1971ApJ...170..265S}
\begin{equation} \label{eq:growth}
 \Gamma_{\rm CR} = \frac{16\pi^2v_{\rm A}}{3\mathcal{F}B_0^2}\left|p^4c\frac{\partial f}{\partial z}\right|_{p=qB_0/(kc)} {.}
\end{equation}
For the damping rate $\Gamma_{\rm D}$, we consider the damping term which is equivalent to cascade $\Gamma_{\rm cas} = (2C_k)^{-3/2}kv_{\rm A}\sqrt{\mathcal{F}(k,z,t)}$ \citep[where $C_k\approx 3.6$,][]{2003A&A...403....1P}. As can be seen, both $\Gamma_{\rm CR}$ and $\Gamma_{\rm D}$ is proportional to $v_{\rm A}$ and thus is proportional to $n_{\rm i}^{-1/2}$, which means the ISM density can affect the growth and damping (mainly growth {here}, i.e., $\Gamma_{\rm CR}>\Gamma_{\rm D}$, because we are discussing the turbulence near the source) of the Alfv\'{e}nic turbulence. For the boundary conditions, we set $\mathcal{F}|_{z=L_{\rm c}/2} = \mathcal{F}_{\rm Gal}$ and $f|_{z=L_{\rm c}/2} = 0$, where $\mathcal{F}_{\rm Gal} = r_{\rm L}c/[3D_{\rm Gal}(E)]$, and $D_{\rm Gal}(E) = 3.6\times10^{28}E^{1/3}_{\rm GeV}$ cm$^2$ s$^{-1}$ \citep{2009APh....31..284P}.

\citet{2021ApJ...919...32B} considered the situation in which only {the} CRs with the highest energy can escape due to the turbulence generated via non-resonant branch of streaming instability. Here we consider another situation in which the accelerated CRs are released into the ISM all at once. Then, the injection is described as $Q(p,z,t) = Q_1p^{-\alpha}e^{-p/p_{\rm cut}}\delta(z)\delta(t)$, where $p_{\rm cut}$ represents the cutoff energy \citep[$\sim 100$ TeV,][]{2013MNRAS.431..415B}, $\alpha$ the spectral index (=4 for canonical diffusive shock acceleration), {and} $Q_1$ the injection constant which can be determined by the total energy of the CRs $E_{\rm CR, tot}$, i.e.,
\begin{equation}
\begin{split}
E_{\rm CR, tot} &= \pi R^2_{\rm inj}Q_1\\
&\times\int_0^\infty  4\pi p^2 \left(\sqrt{c^2p^2+m_{\rm p}^2c^4}-m_{\rm p}c^2\right) p^{-\alpha}e^{-p/p_{\rm cut}}\,{\rm d}p
\end{split}
\end{equation}
{with} $R_{\rm inj}$ the injection scale of the CRs (comparable to the radius of the SNR).

Prior to commencing complicated numerical calculations, it is worthwhile to get some hints from the analytical methods. As is shown in \citet{2008AdSpR..42..486P}, in fully ionized plasma, if we drop the advection and adiabatic terms in \autoref{eq:dif}, the analytical solution of the diffusion coefficient reads \citep{2008AdSpR..42..486P}
\begin{equation}\label{Eq:D}
D(p,z,t) = \frac{2^2\Gamma^8(1/4)\kappa^6t^5}{3^5\Gamma^4(1/2)\eta^4z^2} + \frac{3z^2}{4t},
\end{equation}
{where
\begin{equation}\label{Eq:8}
\kappa = \frac{(cr_{\rm g})^{1/3}B_0^{4/3}}{2^{5/3}3^{1/3}\pi^{7/3}C_{\rm K}p^{8/3}}
\end{equation}
(with $r_{\rm g}$ being the Lamour radius of the protons) and $\eta = \int_{-\infty}^{\infty} f {\rm d} z$ 
},
and the inverse of $f$ reads \citep{{2008AdSpR..42..486P}}
\begin{equation}\label{Eq:9}
f^{-1}(p,z,t) = \sqrt{\frac{\Gamma^8(1/4)}{27\Gamma^4(1/2)}\frac{\kappa^3t^3}{\eta^4} +\frac{27}{16}\frac{z^4}{\kappa^3t^3}}.
\end{equation}

Obviously, when $z^2/t$ is large enough ({$\gg(2^4/3^6)\Gamma^8(1/4)\Gamma^{-4}(1/2) \kappa^6t^5/z^2$;} in other words, the CR energy is low enough), $D$ becomes energy-independent. {For example, for canonical values} $E_{\rm CR, tot} = 10^{50}$\,erg and $\alpha=4$, %{$R_{\rm inj} = 2\,{\rm pc}$, $p=10^6m_{\rm p}c$, $t=3$\,kyr and $z = 5$ pc,}
{
\begin{equation}\nonumber
%\begin{split}
D(p,z,t)  \approx 2.2 \times 10^{20} \left(\frac{R_{\rm inj}}{2\,{\rm pc}}\right)^8 \left(\frac{p}{10^6\,m_{\rm p}c}\right)^2 \left(\frac{t}{\rm 1\ kyr}\right)^5\left(\frac{z}{1\rm \ pc}\right)^{-2}
\end{equation}
\begin{equation}\label{eq:fm1}
 \hspace{1.5cm}+ 2.3\times 10^{26}\left(\frac{z}{1\,{\rm pc}}\right)^2 \left(\frac{t}{1\,{\rm kyr}}\right)^{-1}{\rm cm}^2\ {\rm s}^{-1}.
%\end{split}
\end{equation}}
%\begin{equation}\label{eq:fm1}
%\begin{split}
%{
%D(p,z,t) \approx 2.1 \times 10^{19} {\rm cm}^2{\rm s}^{-1} + \frac{3z^2}{4t}.
}
%\end{split}
%\end{equation}

%Then the CR spectrum can be evaluated via the balance between the turbulence growth and damping $\Gamma_{\rm CR} = \Gamma_{\rm cas}$, 

Then, {we obtain from Equations \ref{Eq:9} and \ref{Eq:8}}
\begin{equation}
{f(p,z,t)  \propto \sqrt{\frac{\kappa^3t^3}{z^4}}} \propto p^{-3.5}.
\end{equation}
That is, independent of the power-law index of injected CR spectrum $\alpha$, at the location in which $z^2/t$ is large enough, the energy spectrum of CR will be as hard as ${\rm d}N/{\rm d}E \propto E^{-1.5}$. {In \autoref{fig:fR1}, we compare the analytical solution and the numerical solution of $D$ and $f$. As can be seen, the numerical solution roughly goes as $f(p) \propto p^{-3.5}$ as is expected, and the diffusion coefficient is almost energy-independent before the escape dominates. In \autoref{fig:fR2}, we also plot the spatial and energy dependence of particle distribution function for reference. As can be seen from the left panel, most of the particles are confined at $\sim 2$ pc from the injection position. In the right panel, it can be seen that 1 PeV particles have partially escaped from the system, while the relatively low-energy particles are still confined well.}

\subsection{Application to G106.3+2.7}\label{sec:3}
We use the code ``AAfragpy'' \citep{2021PhRvD.104l3027K} to calculate the pionic $\gamma-$ray flux. The absorption of $\gamma$-rays during propagation through the $\gamma$-$\gamma$ pair production on CMB photons is considered. The MC is modeled as a cylinder starting at $z=z_1$ and ending at $z=z_2$ saturating the {magnetic flux} tube, with a total mass of $M_{\rm MC}$. We also consider the ISM within the magnetic flux tube as another target for the $pp$ collision. 

In addition to the ISM density, there are other parameters that can affect the predicted $\gamma$-ray flux. To study the effect of the ISM density, we first model the $\gamma$-ray emission in a reference scenario, and then examine the influence of some parameters on the allowable range of the ISM density. 
{The reference scenario uses intermediate-level parameters to establish a baseline, enabling the subsequent investigation of the influence of individual parameters.}
It is worth noting that the MAGIC data and the LHAASO/{Tibet}\,AS$\gamma$ data are somewhat inconsistent {with each other}. As is discussed by \citet{2023A&A...671A..12M}, the MAGIC data may miss some $\gamma$-ray emission in the tail region. On the other hand, HAWC, Tibet AS$\gamma$ and LHAASO data may contain {the} emissions from both the head region and the tail region. As suggested by some previous studies \citep[e.g.][]{2022Univ....8..547L}, the Boomerang Nebula might partially contribute to $\gamma$-ray flux above several hundreds of TeV. Therefore, we will fit the GeV flux, which is concentrated at the tail region of the SNR, observed by Fermi-LAT, and the flux measured by LHAASO up to 100\,TeV. {However,} we consider the flux above 100\,TeV measured by LHAASO as an upper limit for the SNR's emission. 

In \autoref{fig:fitting}, we show the reference scenario and the parameters used to fit the data are listed in \autoref{tab:par}. We explore the dependence of {the} $\gamma$-ray spectr{um} upon the ISM density as is shown in \autoref{fig:density}. The ISM density can affect the GeV $\gamma$-ray spectrum, because the bulk of low-energy GeV protons have not arrived at the MC located at 20--30\,pc away within the age of the SNR, and most of GeV emissions are produced by interactions with the intervening ISM between the source and the MC. Besides, the ISM density can also affect the resulting high-energy cutoff energy, since the growth rate (\autoref{eq:growth}, and the effective growth rate $\Gamma_{\rm CR}-\Gamma_{\rm D}$) is proportional to the Alfv\'{e}n speed. The increase in the ISM density will reduce the Alfv{\'e}n speed and thus the growth of the turbulence will decrease correspondingly. As a consequence, it would aggravate the escape of the high-energy particles from the system before the balance {between the growth and the damping of the turbulence} can be built. {Note that we do not aim to fit all the GeV-PeV $\gamma$-ray data with the model, but rather use them as the upper limits to constrain the ISM density. This is because the GeV flux as measured by Fermi-LAT may be partially contributed by the IC radiation of electrons. If so, the hadronic $\gamma$-ray flux need be adjusted to a lower level in order not to overshoot the measured flux. On the other hand, the $\gamma$-ray flux above 100\,TeV may partially originate from the Boomerang PWN \citep{2022Univ....8..547L, 2020ApJ...897L..34L}.
As shown in \autoref{fig:density}, we can see that an ISM density in the range of $n_{\rm i}=0.01$--$0.05$\,cm$^{-3}$ are consistent with the data in the reference scenario. A higher ISM density will lead to an overshoot in the GeV $\gamma$-ray flux, while a lower ISM density would make a too high theoretical flux compared to the data around 100\,TeV and above. However, the excess in the high-energy end may be avoided if the maximum proton acceleration energy is assumed to be smaller. So the constraint mainly poses an upper limit of $0.05\,\rm cm^{-3}$ for the ISM density, supporting the picture that the SNR breaks into a low-density cavity, as revealed by the X-ray observation.
}

The constraints on the ISM density from the $\gamma$-ray data are also relevant with some other model parameters, which are not necessarily equal to the values in the reference scenario.
In \autoref{fig:various}, we show the $\gamma$-ray spectra obtained {by} setting different value{s} of the coherence length of the interstellar magnetic field (the top-left panel), the age of the SNR (the top-right panel), the position of the MC (the middle-left panel), the spectral index of injected protons (the middle-right panel), and the radius of the tube (the bottom-left panel), other than those in the reference scenario. {Under each of the variation scenario, we then increase the ISM density until the resulting $\gamma$-ray flux reaches the 95\% confidence level upper bound of any spectral data point of Fermi-LAT. As such, we obtain an upper limit of the ISM density in each scenario. } As can be seen from \autoref{fig:various} and will be discussed below, {our results do not vary significantly with these parameters, as long as the choice of parameters is reasonable. Note that, effects of changing different parameters may also cancel each other. So uncertainties of other parameters basically do not affect our conclusion that the tail of the SNR is expanding in a low-density cavity. }

The coherence length determines the escape timescale of the particles, and therefore affects the cutoff energy of the $\gamma$-ray spectra. Meanwhile, the SNR age will also affect the cutoff energy of the spectra as the high-energy particles will escape gradually over time. The influence of the increased/decreased SNR age can be compensated by the increased/decreased coherence length. {Also, the excess in the high-energy spectral end can be always cancelled by introducing a smaller cutoff energy at the injection spectrum.} These two parameters almost do not have influence on the low-energy $\gamma$-ray flux and hence the constraint on the ISM density from the GeV data is unaffected. 

The position of the MC will also affect the $\gamma$-ray spectra. The larger distance {from the SNR/release site} will lead to a harder $\gamma$-ray spectra, because a smaller fraction of low-energy particles can reach the MC within the SNR age. As shown in \autoref{fig:fR2}, the spatial distribution of particles are highly energy-dependent. {As the distance increases, the proton spectra becomes harder and the low-energy gamma-ray flux decreases. An upper limit of $n=0.1\rm cm^{-3}$ for the ISM density can be obtained in the case of $z_1=36\,$pc. As can be seen, the SNR-MC distance will not have much effect on the density.} 

The injection power-law index determines the fraction of energy occupied by high-energy particles. While the index is not supposed to deviate from 2 too much according to the canonical shock acceleration theory, we here vary it slightly to see the influence. The resonant branch of the streaming instability is determined by the number density and number density gradient of high-energy particles. Therefore, a softer injection spectrum will lead to fewer high-energy particles and, consequently, weaker generated turbulence resonating with high-energy particles. The escape timescale of high-energy particles would become shorter remarkably for a softer injection spectrum. On the low-energy side, although the total number of particle{s} (i.e., $\eta$) would increase significantly with a softer injection spectrum, {self-regulation of particles can be established more quickly than in the case of a harder injection spectrum.} The second term in the right hand side of \autoref{eq:fm1} dominates and most of the low-energy particles are confined within a small distance ($\sim 1$ pc) from the injection position, due to self-similarity of the system. Therefore, the increase in low-energy particles is not substantial at large distance. {In the case of $\alpha=3.8$ and $4.2$, the upper limits of ISM density are found to be $0.02\rm cm^{-3}$ and $0.1\rm cm^{-3}$, respectively.}

%Considering the fact that a fraction of 100\,TeV $\gamma$-rays may come from the PWN, a larger tube size leads to a larger ISM density. {Consider the physical size of SNR G106.3+2.7, the ISM density cannot exceed 0.1 cm$^{-3}$.}
The radius of the magnetic tube has a systematic influence on the energy density and the energy density gradient of high-energy particles. A larger tube radius would increase the volume of the tube and hence dilute the particle density, and vice versa. As a result, for the smaller tube size the cutoff energy is higher, because a smaller tube size will lead to a higher particle density and density gradient (in order to have the same total particle number). {As is shown in \autoref{fig:fR1}, the numerical solution will deviate from the analytical solution as the energy increases. This deviation occurs because the particle number density and density gradient are not large enough to reach self-regulation at high energies, and these particles diffuse more quickly than that predicted by the analytical solution (which assumes the self-regulation has been built). Hence, {a larger tube size will relax the constraint on the ISM density. We find that with a tube radius of 4\,pc, the upper limit of the ISM density is increased to 0.3\,$\rm cm^{-3}$. Although this upper limit is 6 times higher than the one obtained in the reference scenario, it is still consistent with the picture of a low-density cavity around the tail region of the SNR. In fact, the ISM density derived from the X-ray observation may be also used as a constraint on tube radius in this case.} 

{We also explore the difference between an impulsive injection at $t=0$ and a constant injection of CRs. If the SNR has not been significantly decelerated, it is probably still processing as a proton PeVatron, and hence we may expect the particle injection is continuous. For simplicity, we assume a constant particle injection rate. In this case, more high-energy particles remain in the tube compared to the impulsive injection case, as more particles are injected recently and do not have time to escape the tube. Therefore, the resulting $\gamma$-ray spectra at the MC show a larger cutoff energy. On the other hand, the density of low-energy particles at the MC is not significantly affected as self-regulation is established at early time. Thus, the obtained constraint on the ISM density remains unchanged even if the particle injection rate is constant.}

Finally, we discuss the total amount of the injected proton energy. {In the above calculation, we fix the total CR energy to $4\times 10^{49}\,$ergs which is relatively small compared with the typical proton energy budget of an SN explosion, i.e., a few times $10^{50}$\,erg.} In general, it is believed that the SNR blast wave can convert $\sim 10\%$ of its kinetic energy to {the} accelerated nonthermal particles, so the total injected proton energy may have an uncertainty of a factor of a few. From the perspective of $\gamma$-ray production, the amount of total proton energy is in degeneracy with the target gas density. However, the influence is not linear when considering the self-regulated transport of injected protons. A smaller injection normalisation will lead to less particle density gradient in weaker turbulence {at high energies}. It would then lead to an insufficient amount of high-energy protons at the MC to explain the multi-TeV data, and this deficiency cannot be compensated by employing a higher MC mass which is estimated in the range of 300 -- 800 $M_\odot$ \citep{2022ApJ...926..124L}. On the contrary, if we adopt a larger total proton energy, it would only lead to a more stringent constraint on the ISM density (i.e., a lower upper limit of $n_i$).

{
\subsection{Farmer-Goldreich damping and ion-neutral damping}
Here we discuss Farmer-Goldreich damping \citep{2004ApJ...604..671F,2016ApJ...833..131L} and ion-neutral damping. Subsequent analysis will demonstrate that the Farmer-Goldreich damping is negligible. Similarly, we will show that ion-neutral damping is also negligible. The damping arise from interactions with pre-existing turbulence can be written as 
\begin{equation}
\Gamma_{\rm FG} = 2 \frac{v_{\rm A}M_{\rm A}^2}{r_{\rm L}^{1/2}L_{\rm MHD}^{1/2}},
\end{equation}
where $M_{\rm A}$ is the Alfv\'{e}n Mach number, and $L_{\rm MHD}$ the injection scale. The damping rate $\Gamma_{\rm IND}$ is derived by solving the equation outlined in \citep{1982ApJ...259..859Z}.
\begin{equation}\label{eq:IND0}
\omega(\omega^2-\omega_k^2)+i\nu \left[(1+\epsilon)\omega^2-\epsilon\omega_k^2 \right]=0,
\end{equation} 
where $\omega$ is the complex frequency of the wave, $\omega_k = kB_0/\sqrt{4\pi n_{\rm i}m_{\rm i}}$, $\nu$ ion-neutral momentum transfer ratio 
\begin{equation}
\nu \approx 1.68 \times 10^{-8}\left(\frac{n_{\rm n}}{{\rm cm}^{-3}}\right) \left(\frac{T}{10^4{\rm\,K}}\right)^{0.4},
\end{equation}
and $\epsilon=n_{\rm i}/n_{\rm n}$ is the ratio between the densities of ions and neutrals. With $\omega = \omega_{\rm R} + i\Gamma_{\rm IND}/2$, \autoref{eq:IND0} can be written as 
\begin{equation}
\omega_{k}^2 = -\frac{\Gamma_{\rm IND}}{\Gamma_{\rm IND}+\nu}\left[\Gamma_{\rm IND}+\nu(1+\epsilon)\right]^2.    
\end{equation}

X-ray observations have uncovered that the cavity density is remarkably low. Correspondingly, the environment's temperature is expected to be exceptionally high, leading to near-complete ionization. Due to this low density, the gas's cooling timescale exceeds 1 Myr \citep[exceeding $10^6$ K,][]{1998ppim.book.....S}. Additionally, the pressure balance also requires a high temperature, as the nearby denser gas should has density of $\sim 1$\,cm$^{-3}$ and temperature of $10^4$ K. The ionization rate can be estimated via Saha equation. The result is that, under $10^6$\,K, the hydrogen almost fully ionized, and about 90\% of the helium are also ionized, i.e., the neutral fraction is $\sim 1\%$. 

In \autoref{fig:INDFG}, we show the difference in whether considering $\Gamma_{\rm FG}$ (with an unity $M_{\rm A}$ and $L_{\rm MHD}=L_{\rm c}$) and $\Gamma_{\rm IND}$. As can be seen, both damping can be neglected. In considering IND, $n_{\rm n}=0.0003$, $n_{\rm i}= 0.0297$, $T=10^6$ K. As can be seen, the difference is negligible.

}

\section{discussion}\label{sec:5}
%In this paper, we only consider the resonant branch of streaming instability. The non-resonant Bell mode \citep{2004MNRAS.353..550B} is suggested to grow much faster than the resonant modes when the current of escaped CRs is large enough, and is significant in confining the higher-energy CRs whose density (and density gradient) is lower than lower-energy ones. Although the resonant branch works well here, there still exists probability that the high energy CRs are scattered by the turbulence driven by Bell mode.

The low ISM density indicates that the swept-up mass by the SNR blast wave is $\sim 0.1\,M_\odot$. With such a low swept mass, the SNR can hardly enter the Sedov-Taylor phase at its current age and the shock is not severely decelerated. In this case, the shock velocity can simply be obtained via dividing the SNR size by the SNR age. If we consider {the} current position of the pulsar as the explosion position, the shock velocity is {about} $\sim 6000(d/800\,{\rm pc})(t_{\rm SNR}/1\,{\rm kyr})\,$km/s, which is consistent with the X-ray observation and the fact that PeV particle{s have} been accelerated. In this paper, we constrain the ISM density in the tail region of the SNR considering an asymmetric expanding shock. In \citet{2021ApJ...919...32B} and \citet{2022A&A...658A..60Y}, larger densities (0.2\,cm$^{-3}$ and 0.5\,cm$^{-3}$) are used to explain the size of the SNR with an isotropic expanding model{. T}he larger density arises from the fact that there is a denser HI region in the northeast ``head'' region as is suggested by \citet{2001ApJ...560..236K}, and the density is averaged over 4$\pi$ solid angle. While the tail region is still in the free expansion phase, the head region should have already entered the Sedov phase, and can no longer accelerate PeV protons. Although in the first hundred years the shock in head region could accelerate protons to PeV, these protons may have escaped from the shock already and are diffusing in the ambient ISM. It may suggest that the 100\,TeV $\gamma$-rays \citep{2021Natur.594...33C} seen in the head region stems from the pulsar. 

{Young SNRs {such as} SN 1006, RXJ 1713, and Vela Jr., characterised by fast shock velocities, typically exhibit bright synchrotron X-ray filaments amplified by strong magnetic fields \citep[see e.g.,][]{2014ApJ...790...85R}. However, filaments are absent in SNR G106.3+2.7. The explanation is that, for perpendicular shocks, electron injection is hard to trigger \citep{2023PPCF...65a4002B}, and a quasi-perpendicular SNR shock may form for SNRs expanding in stellar wind cavities \citep{1988ApJ...333L..65V,2018APh....98...21Z}. So even if the magnetic field is enhanced, {it may also be hard} to see the filaments for perpendicular shock because there are not enough accelerated electrons.}

%，{Since Tibet AS$\gamma$ and LHAASO data below 100 TeV may contain emissions from both the head region and the tail region, the pure hadronic scenario is only possibility among other scenarios. However, the conclusion of this paper is still robust, as the upper limit mainly comes from the GeV fluxes, and the density could be even lower if GeV emissions in tail region are partly contributed by electrons accelerated by the PWN.}

\section{conclusion}\label{sec:4}
In this paper, we {constrain} the ISM density around the PeVatron candidate, SNR G106.3+2.7, with the X-ray and $\gamma$-ray observations. We {develop} a model for the self-regulated propagation of CR protons injected from the SNR and {calculate} the $\gamma$-ray emission of {the} CRs via hadronuclear interactions with the {MC} and the ISM. We found that non-detection of the thermal X-ray emission from the tail region may impose an upper limit of $0.02-0.07\,\rm cm^{-3}$ for the ISM density. The measurement of GeV-PeV $\gamma$-ray flux may also constrain the density of the ISM where the SNR is expanding. 

Our calculations {demonstrate} that, apart from the influence on the X-ray emission, the ISM density also plays a crucial role in shaping the GeV $\gamma$-ray flux and the high-energy cutoff of the resulting {$\gamma$}-ray spectrum. In order to achieve an acceptable fit to the GeV-PeV $\gamma$-ray data and {a reconciliation} with {the} X-ray observations, the ISM density should lie $\sim 0.01-0.05$\,cm$^{-3}$ for the reference scenario. {Taking the observed $\gamma$-ray data as upper limits of the hadronic emission from the CR interactions with the ambient medium, we {find} that most of the other parameters have minor impact on the constraint of the ISM density up to a factor of a few}. Our results is in agreement with the scenario that the SNR is expanding into a low-density cavity, which may enable it to be a proton PeVatron.
%%%%%%%%%%%%%%%%%%%%%%%%%%%%%%%%%%%%%%%%%%%%%%%%%%
\section*{Acknowledgements}
We thank Lei Sun and Wei Sun for helpful discussion about X-ray observation. This work is supported by 
NSFC under grant No. 12173018, 12121003, U2031105, and U1931204. 

\section*{Data Availability}
The data underlying this article will be shared on reasonable request to the corresponding author.

\begin{center}
\begin{figure*}
\includegraphics[scale=0.5]{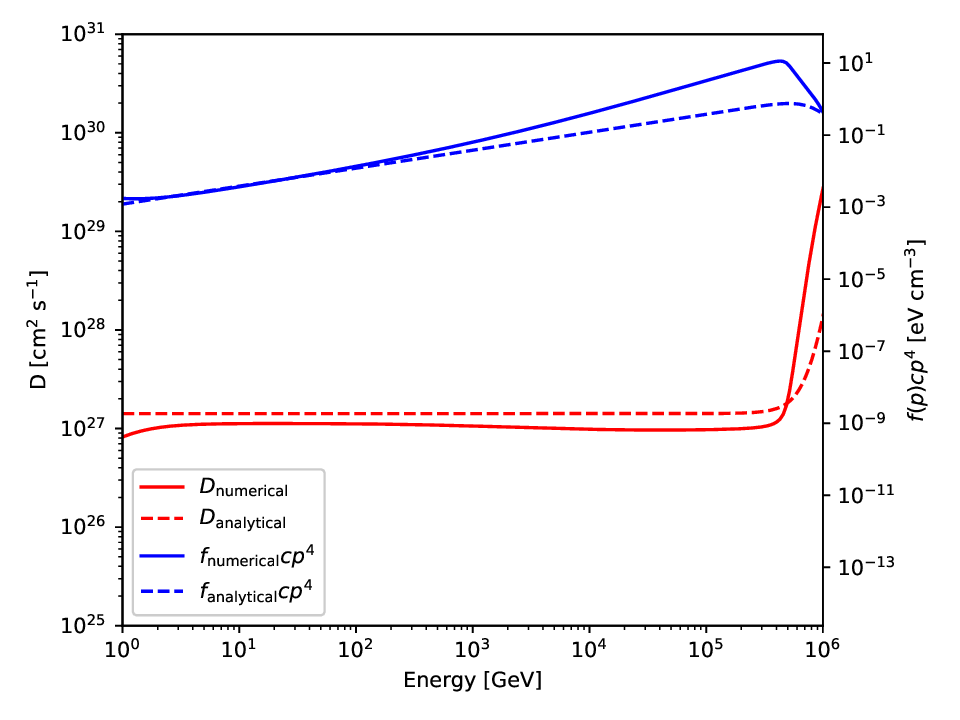}
	\caption{The diffusion coefficients and particle distribution function at $t=4$ kyr and $z=5$ pc. }
\label{fig:fR1}
\end{figure*}
\end{center}

\begin{center}
\begin{figure*}
\includegraphics[scale=0.5]{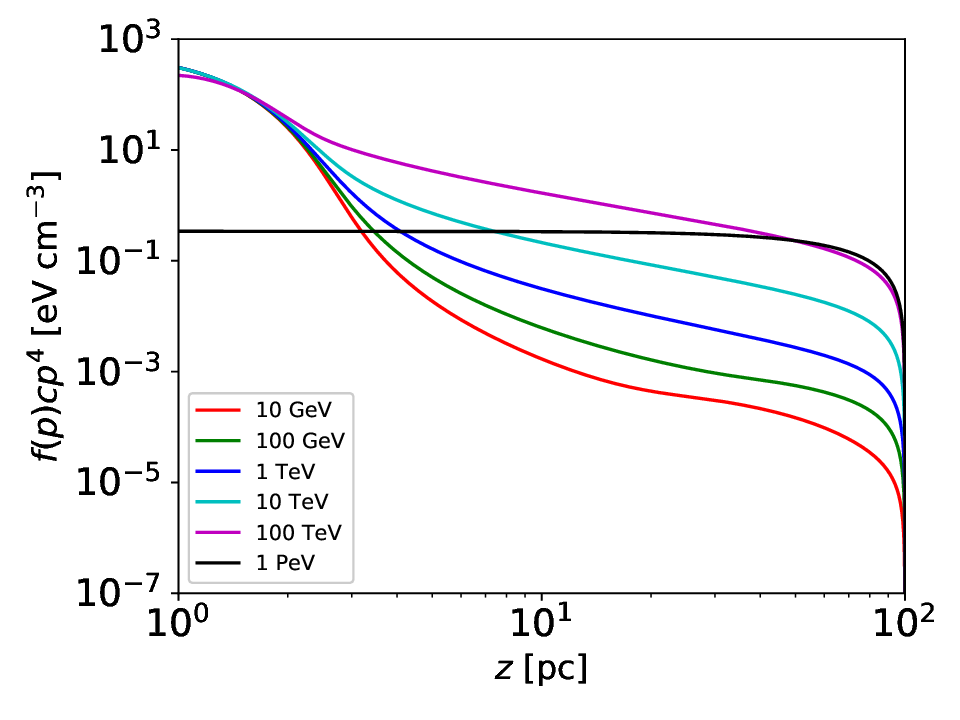}
\includegraphics[scale=0.5]{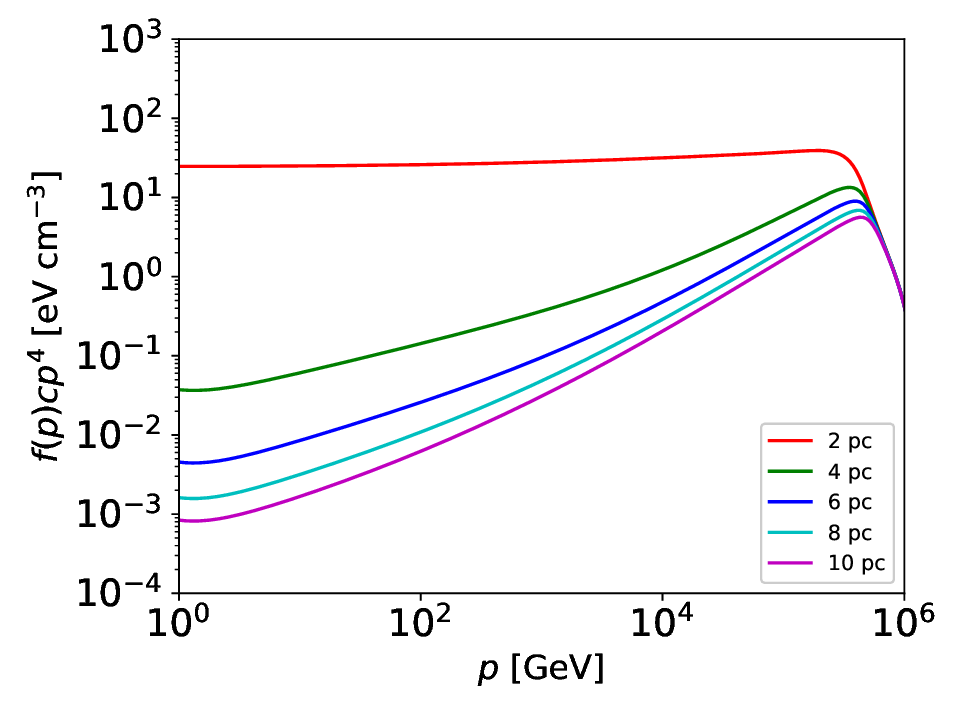}
	\caption{Dependence of particle distribution function $f(p)$ upon distance and energy.}
\label{fig:fR2}
\end{figure*}
\end{center}

\begin{center}
\begin{figure*}
\includegraphics[scale=0.5]{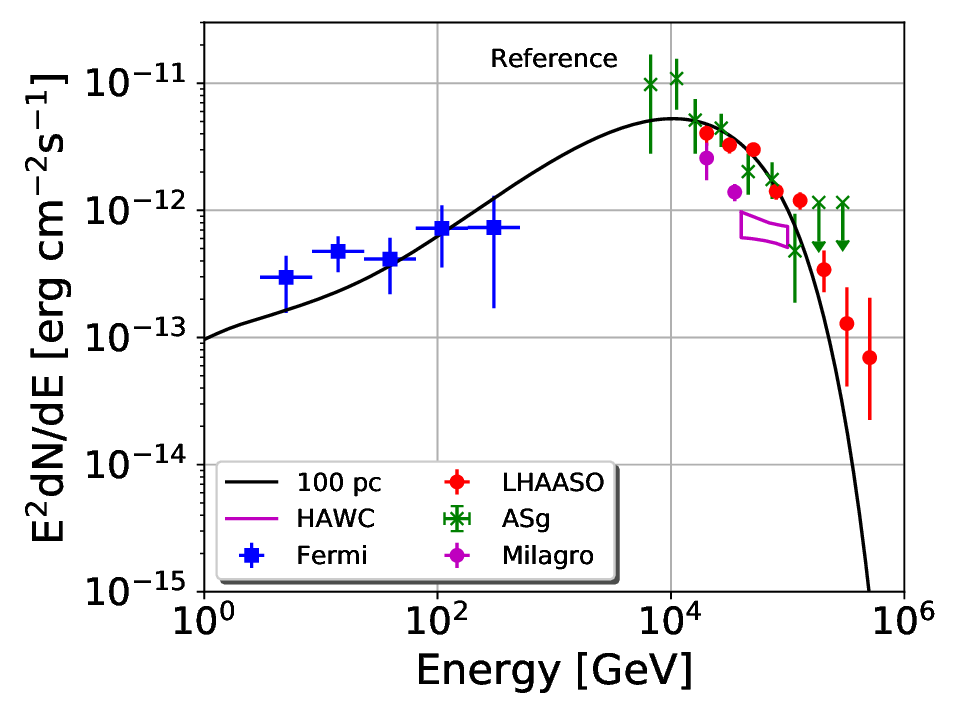}
	\caption{$\gamma$-ray spectra of SNR G106.3+2.7. The Fermi data are taken from \citep{2019ApJ...885..162X,2020ApJ...897L..34L}, HAWC$\gamma$ data from \citet{2020ApJ...896L..29A}, Tibet AS$\gamma$ data from \citet{2021NatAs...5..460T}, and LHAASO data from \citet{2021Natur.594...33C}.}
\label{fig:fitting}
\end{figure*}
\end{center}

\begin{center}
\begin{figure*}
\includegraphics[scale=0.5]{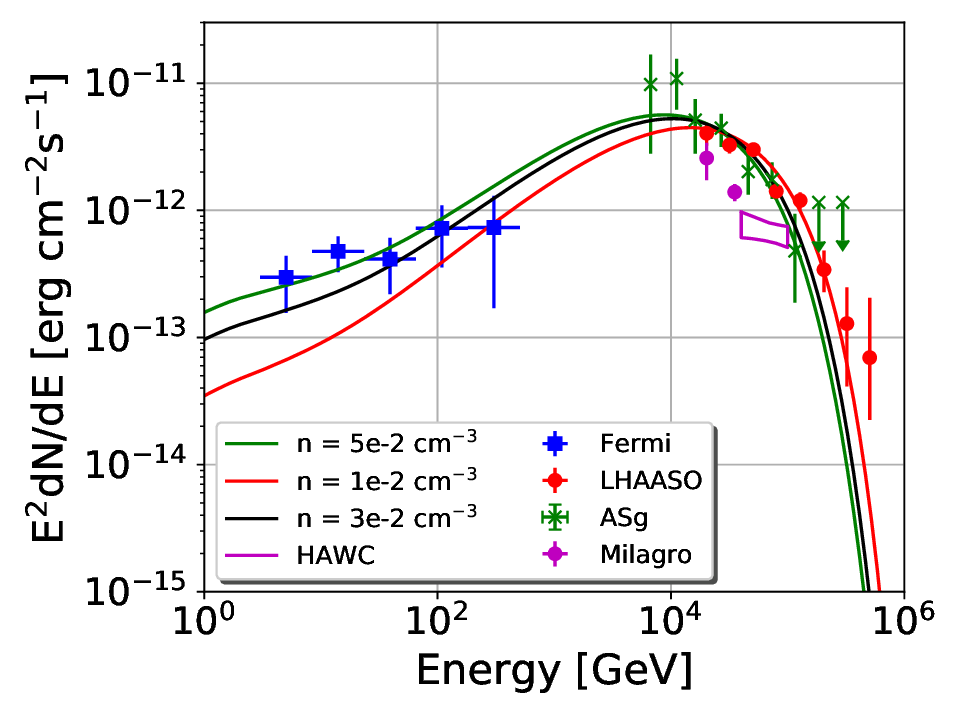}
	\caption{The dependence of $\gamma$-ray spectra of SNR G106.3+2.7 on density.}
\label{fig:density}
\end{figure*}
\end{center}

\begin{center}
\begin{figure*}
\includegraphics[scale=0.5]{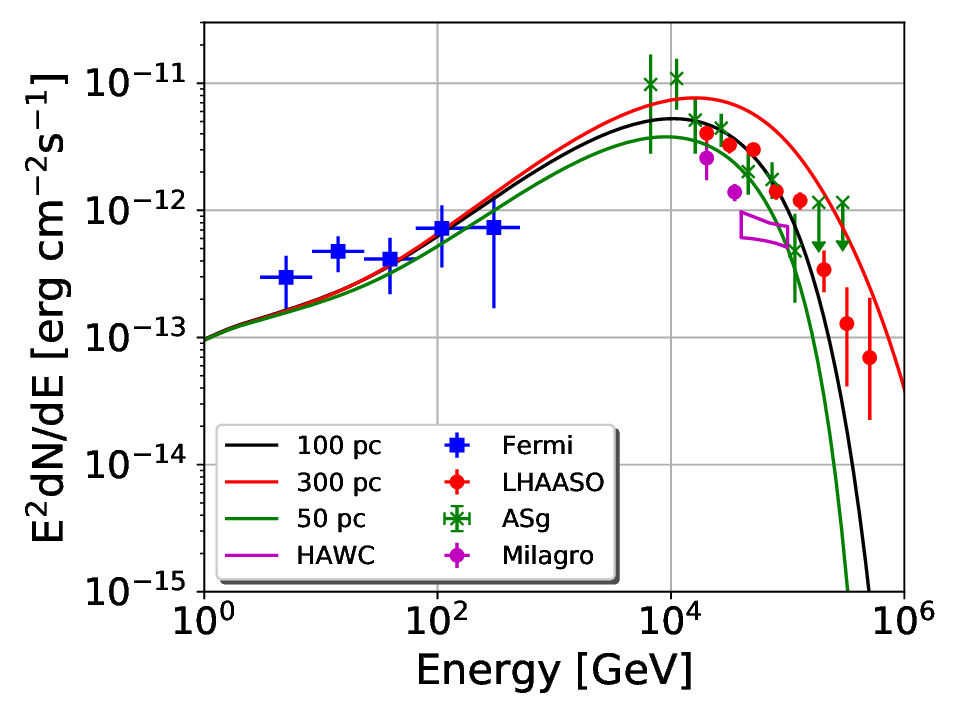}
%100pc & 300pc n<0.06; 50 pc n<0.09
\includegraphics[scale=0.5]{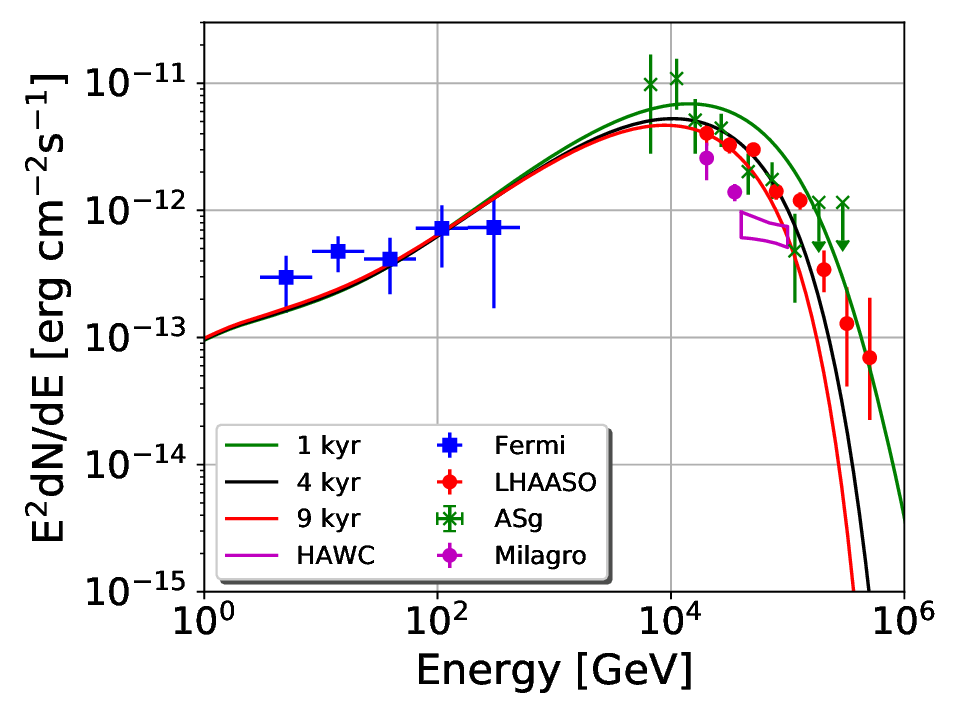}\\
% n<0.06
\includegraphics[scale=0.5]{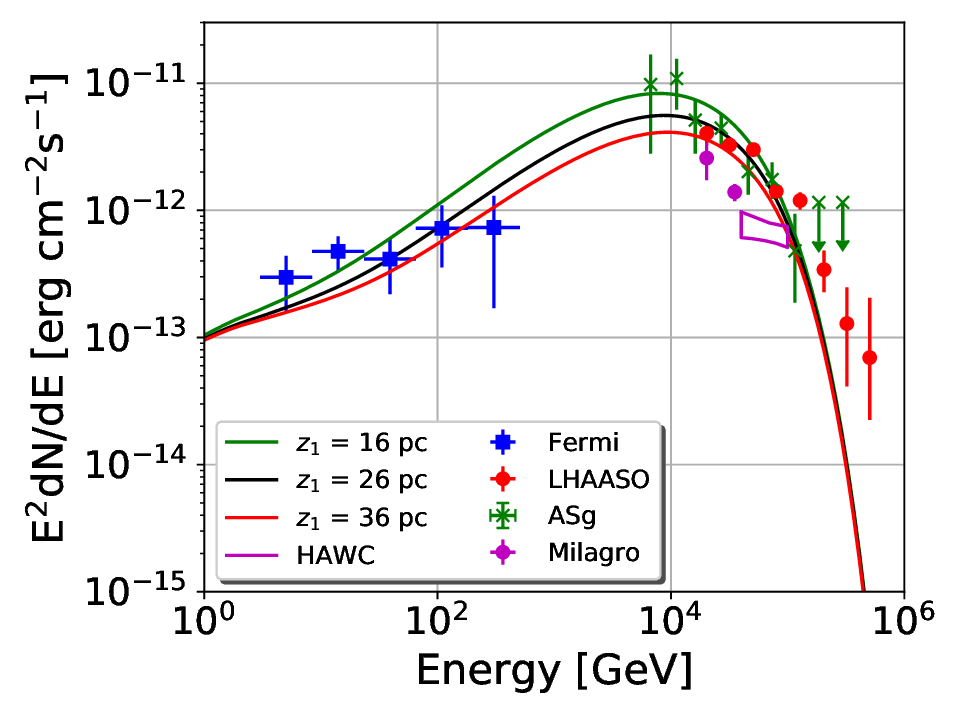}
%z=16 n<0.02; 36pc n< 0.1
\includegraphics[scale=0.5]{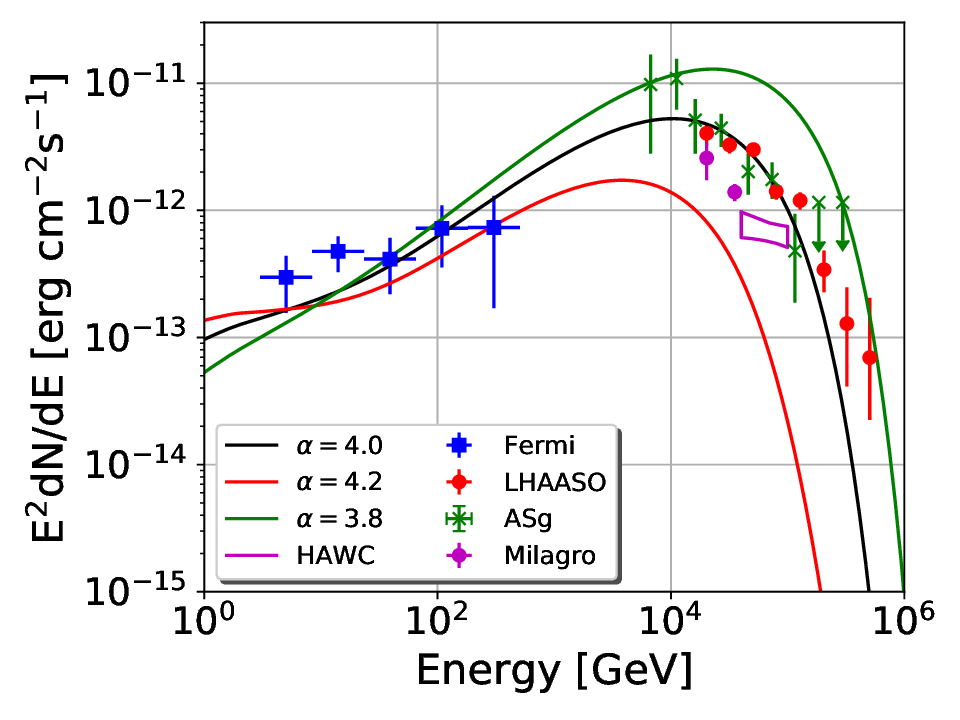}\\
%alpha=4.2, n<0.105, alpha=3.8, n<0.025
\includegraphics[scale=0.5]{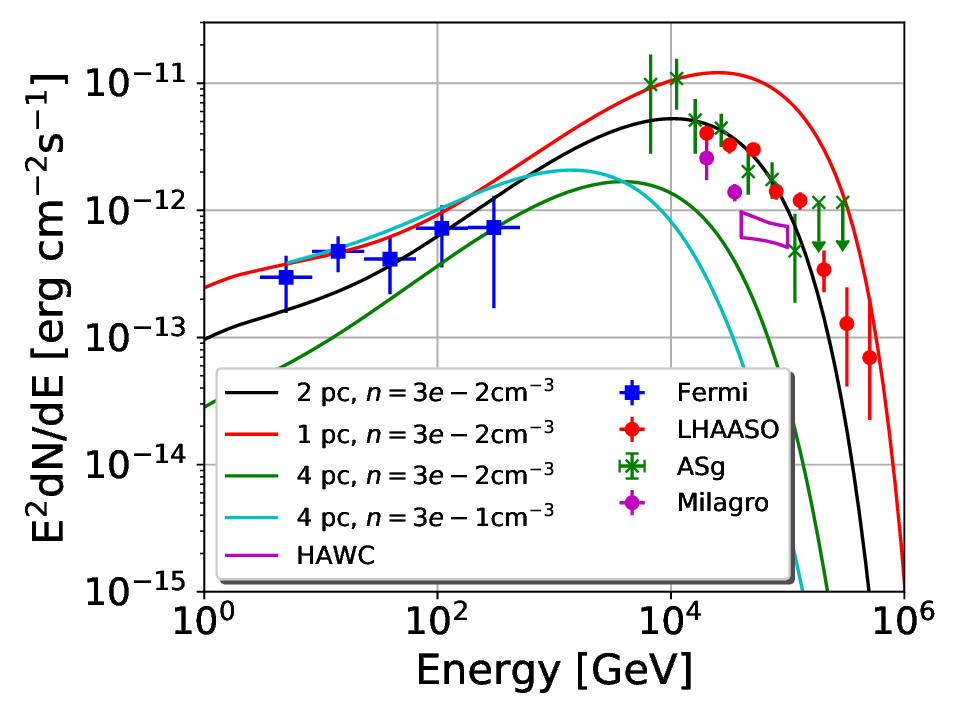}
%tube=1pc, n< 0.019; tube = 4pc, n<0.35
\includegraphics[scale=0.5]{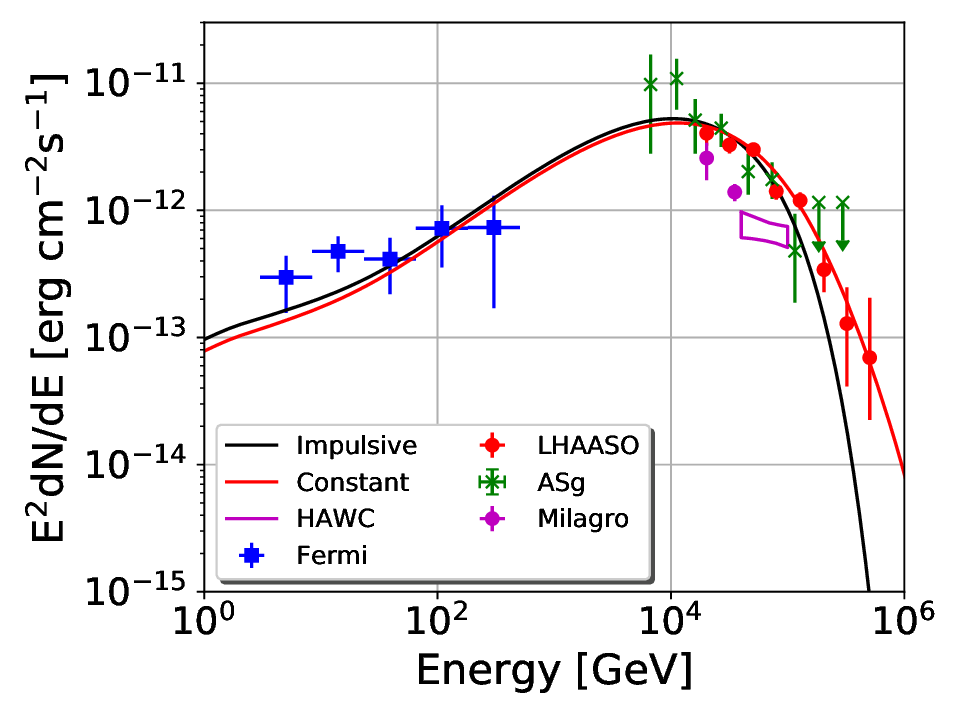}
% n<0.06

	\caption{Dependence of $\gamma$-ray spectra on coherence length (upper left), age of the SNR (upper right), MC position (middle left), injection power-law index (middle right), tube size (lower left), injection function (lower right).}
\label{fig:various}
\end{figure*}
\end{center}

\begin{center}
\begin{figure}
\includegraphics[scale=0.5]{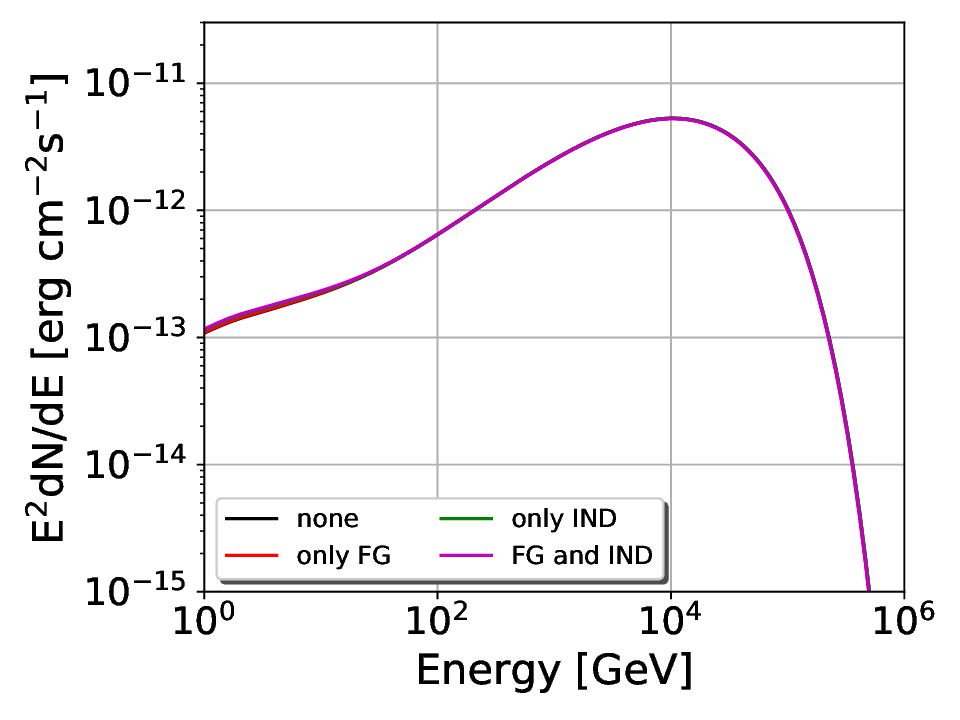}
\caption{SED incorporating only $\Gamma_{\rm FG}$ (green), only $\Gamma_{\rm IND}$ (red), both (magentar) and none (black)}
\label{fig:INDFG}
\end{figure}
\end{center}

\begin{table*}
\caption{Reference Parameters\label{tab:par}}
 \begin{tabular*}{\columnwidth}{l@{\hspace*{40pt}}l@{\hspace*{40pt}}l}
 \hline\hline
Parameter     & Value\\
  \hline
$T_{\rm age}$ (kyr)                       	& 4.0 \\
$p_{\rm cut}$ (PeV)					        & 1.0\\
$\alpha$ 									& 4.0\\
$R_{\rm inj}$ (pc)                          & {2.0}\\
$E_{\rm CR, tot}$ ($10^{50}$ erg)		    & {0.4}\\
$d$ (kpc)                       			& {0.8}\\
$n_{\rm i}$ (cm$^{-3}$)					    & {0.03}\\
$n_{\rm n}$  (cm$^{-3}$)	            	& {0}\\
$B_0$ ($\mu$G)								& {3.0}\\
\hline
\hline
$z_1$ (pc)									& 26 \\
$z_2$ (pc)									& 28 \\
%2pc*pi*2pc**2*150 *1.4*1.67e-24
$M_{\rm MC}$ ({$M_\odot$})      & 120 \\
{$L_{\rm c}$ (pc)}              & {100} \\
 \end{tabular*}
\end{table*}



\bibliographystyle{mnras}
\bibliography{example} % if your bibtex file is called example.bib

%\begin{thebibliography}{99}
%\end{thebibliography}

%%%%%%%%%%%%%%%%%%%%%%%%%%%%%%%%%%%%%%%%%%%%%%%%%%

%%%%%%%%%%%%%%%%% APPENDICES %%%%%%%%%%%%%%%%%%%%%

%%%%%%%%%%%%%%%%%%%%%%%%%%%%%%%%%%%%%%%%%%%%%%%%%%

% Don't change these lines
\bsp	% typesetting comment
\label{lastpage}
\end{document}